# Harmonic Generation in Metal-Insulator and Metal-Insulator-Metal Nanostructures

Mallik M. R. Hussain[1], Imad Agha[1,*], Zhengning Gao[2], Domenico de Ceglia[3], Maria A. Vincenti[4], Andrew Sarangan[1], Michael Scalora[5], Parag Banerjee[2,6], Joseph W. Haus[1]

[1]*Department of Electro-Optics and Photonics, University of Dayton, OH - 45469, USA*
[2]*Department of Materials Science & Engineering, University of Central Florida, Orlando, FL - 32816, USA*
[3]*Department of Information Engineering, University of Padova, Padova, Italy*
[4]*Department of Information Engineering, University of Brescia, Brescia, Italy*
[5]*US Army AMRDEC, Charles M. Bowden Research Laboratory, Redstone Arsenal, AL - 35898, USA*
[6]*Department of Mechanical Engineering & Materials Science, Washington University, St. Louis, MO - 63130, USA*
*\*Corresponding authors: jhaus1@udayton.edu, iagha1@udayton.edu*

We report that the second and third harmonic signal reductions with insulator film surface coverage over a gold substrate gives a measure of the electron density in the spill out volume of the insulator, which is dubbed *metal insulator gap states*. For metal-insulator-metal (MIM) structures we observe enhancement saturation and quenching of the third harmonic efficiencies well above the efficiencies for metal-insulator (MI) samples. The measured optical harmonics of scattered light from MI and MIM systems are compared with detailed simulations of the nonlinear interactions including free electron spill out into the insulator, nonlocal and electron quantum tunneling effects. Gold coated substrates are covered with variable thin insulator film thicknesses using atomic layer deposition. Optical harmonics of light scattered from two insulator materials (ZnO and $Al_2O_3$) are measured in our experiments. Based on our simulations we conclude that the observed MIM signal enhancement effects are primarily due to nonlocal phenomena in an electron gas.

## 1. INTRODUCTION

Current understanding of the nonlinear properties of materials on the nanometer scale relies on an initimate connection between the proposed theories and the corresponding experiments. However, it is exceedingly difficult to establish a definitive connection between the two because of subtle, competing effects. Recent studies [1-15] use nonlinear light-matter interactions to expose the essential physics at the sub-nanometer scale, which provides further challenges to our understanding of electron dynamics at that scale. We examine the nonlocal interactions derived from the hydrodynamic model of free electrons [1-6] and the quantum tunneling effects [7-15] and compare their predictions with the experimental observations. In the literature, these effects were used independently to predict the behaviour of probing optical phenomena at nanometer and sub-nanometer length scales. For instance, the sub-nanometter spillout of the electron density at the surface of metals has been exposed recently to have a significant effect on the scattered harmonic light [6, 16]. Meanwhile, quantum tunneling effects were reported as the dominant mechanism in quenching the plasmonic field enhancement in recent experimental papers [17-19]. The conclusions of these studies were based on a phenomenological model for quantum tunneling [10] that compares the predictions with results from time-dependent density functional theory (TDDFT). However, TDDFT has severe computational limitations to simulate larger, more realistic systems. To circumvent this limitation, we employ a quantum tunneling theory that does not have any fit parameters [11, 13, 14] and add nonlocal contributions to the currents for harmonic generation in our simulations; we compare the simulation predictions to our experimental results. The quantum theory is based on *quantitative* comparisons with solid state systems including MIM rectenna systems [20] and it too compares quantitatively with the results from TDDFT [10, 15].

Intense coherent light interacting with conducting free electron systems has given rise to the field of nonlinear plasmonics [21], which provides the required information about electron dynamics for our experiment. Our studies have concentrated on nonlinear harmonic generation in metal-insulator-metal (MIM) nanostructures as a key phenomenon both at the basic and applied levels. On the fundamental level, the second and third harmonic signals from nanometer scale systems are used for studying the optical nonlinearity beyond the classical bulk crystal models [22, 23], where geometric and quantum effects can play a crucial role. Moreover, in a field where multiple, distinct models have been proposed for calculating the efficiency of nonlinear generation, we can test their accuracy by comparing the efficiency of nonlinear harmonic generation in MIMs as a function of disparate parameters, such as pump power, insulator film thickness, metal particle size, incidence angle, etc. On the applied level, there is a growing interest in MIM nanostructures for broadband detection [12], nonlinear imaging, [24] as well as photovoltaics [20]. For these applications, nonlinear generation is either directly employed (imaging), or can be used as a tool for indirectly predicting performance (photovoltaics and photodetectors).

In this paper, we quantify the second and third harmonic signals scattered from a structure based on metal, coated by an ultra-thin atomic-layer deposited (ALD) dielectric, and capped by gold spherical nanoparticles. The advantages of this structure are threefold: a) By using atomic layer deposition, we can develop atomic-level control over the thickness and material properties of the deposited film, b) The spherical nanoparticle geometry allows for field enhancement while simultaneously allowing measurement of a portion of the field harmonics as useful signal, and c) the density of nanoparticles can be controlled by calibrating the particle concentration in solution as well as by spin-coating the particles on the substrate at various speeds. The ultimate aim of this study is to quantify the role of quantum tunneling and metal-induced gap states (MIGS) [16] in harmonic generation from MIM nanostructures. This allows us to directly probe the fundamental physical nature of MIM-based nanostructured devices within the context of the role of surface light-matter interaction; and offer critical pathways for optimization the nonlinear conversion effects.

## 2. BACKGROUND PHYSICS

Our simple experimental geometry illustrated in Figure 1 is an insulator region sandwiched between a gold nanosphere and a gold film. This is the simplest form of a MIM structure that can reveal the physical nature of nanoscale systems when interrogated by laser light. Due to charge build-up at the metal boundaries there is an enhanced electric field in the dielectric gap. However, two physical mechanisms affect the field enhancement, quantum tunneling effects and nonlocal response of the free electrons in gold, both effects are included in our simulations. For a given insulator film thickness there are optically activated quantum tunneling processes that elicit electronic currents as a response. For sub-nanometer gap separations, the AC current reduces the charge build-up and thus quenches the field enhancement in the insulator gap between the nanoparticle and the gold film. MIM tunnel junctions' form current paths with insulator thickness in the nanometer to sub-nanometer range.

The quantum currents are described by linear and nonlinear conductance terms [12, 14, 15]. The conductances have a complex dependence on the insulator gap width $w$, but as $w \to 0$ the quantum conductances saturate. For an AC driving field the quantum conductances are related to linear and nonlinear susceptibilities. Quantum mechanically, in the sub-nanometer range for $w$, the electric field enhancement is limited by the onset of tunneling current and, thus, the insulator region acquires new linear and nonlinear susceptibilities (see Supplement). The quantum-based linear susceptibility is related to the linear quantum conductivity ($\sigma_\omega$) by [14, 15],

$$\chi_q^{(1)}(\omega) = i\sigma_\omega/\omega\varepsilon_0. \tag{1}$$

In Equation 1 $\omega$ is the fundamental photon angular frequency, $\varepsilon_0$ is the free space dielectric permittivity. The nonlinear quantum susceptibilities are small over all thickness values we experimentally studied (see Supplement).

The nonlocal effects are expressed within the hydrodynamic free electron model [2-5]. The linearized dynamical equations contain a pressure gradient term whose coefficient is related to the electron density through the electronic Fermi level; the spatial derivatives of the field constitute the nonlocal effects within the hydrodynamic model.

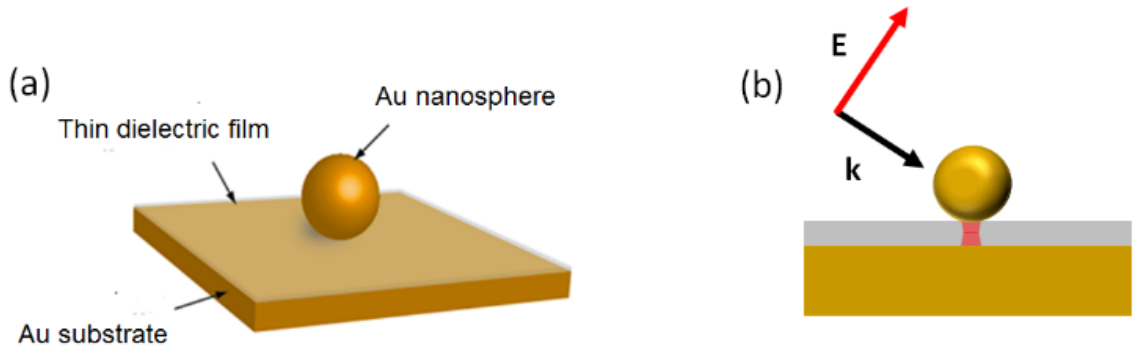

Fig. 1. (a) A gold nano-sphere sits on top of a thin dielectric spacer film deposited on a thick gold film. (b) Side view of the gold nanoparticle with obliquely incident, p-polarized laser light. The field enhancement region in the gap is highlighted in red.

## 3. EXPERIMENTAL PROCEDURE

The experimental setup is schematically illustrated in Figure 2. Our source is a mode-locked Ti:Sapphire laser (Spectra-Physics Tsunami) with 100 fs pulse width at 80 MHz pulse repetition rate. The average output power is 0.5

W and the wavelength is fixed at 810 nm. The pulse passes through a half-wave plate for polarization control and an aspheric focusing lens in the 'input section' before the sample the focused spot size is 25 microns and the peak power is calculated as 6 GW/cm$^2$. The laser beam is incident on the sample surface at an oblique angle of 68º. After the sample, two lenses are used to collimate and refocus the beam; that is followed by two prisms and a short-pass filter placed before the photodetector.

The harmonic light scattered from the sample is separated from the fundamental beam, whose stray light is reduced below measurable limits. The optics forms a demagnified image of the sample surface on the photodetector surface. The demagnification is designed to improve the tolerance of the output section as an imaging system. A TE-cooled UV-enhanced Si detector (EO Systems: UVS-025-TE2-H) is placed at the focal length of L3 to collect the signal. The detector's (PD) output was, then, measured through a lock-in amplifier (LA), SR830. The detector was cooled to $-40^0 C$ and a low gain configuration was used. A lock-in amplifier reduced the noise-equivalent-power below 10fW. The input and output arms were so placed that the specular reflection (the incident angle is kept at 68$^0$) from the sample can be collected.

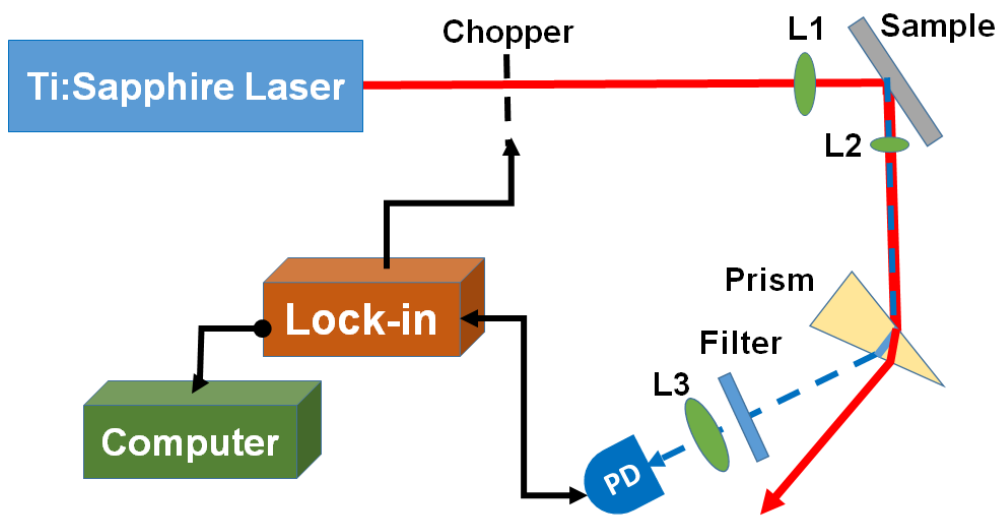

Fig. 2. Illustrated experimental setup: Mode-locked Ti:Sapphire laser, chopper; L1: Focusing lens; L2: collimating lens; filtering prisms; short-pass filter; L3: detector lens; PD: (silicon) photodiode.

The SH and TH signals from the bare Au surface were collected and their efficiencies were compared against the values found in the literature with good agreement. MI samples were fabricated with $Al_2O_3$ or ZnO films of controlled thickness. The thickness change of the insulator film over our gold surfaces (grown by electron beam evaporation on a silicon substrate; surface global roughness ~0.8 nm) required sub-nanometer control. This was achieved using atomic layer deposition (ALD). Details of the ALD process for $Al_2O_3$ can be found in our earlier work [16]; additional information for ZnO deposition can be found in the Supplement.

By varying the number of ALD cycles, metal-insulator (MI) samples were fabricated with different insulator thicknesses (total number of ALD cycle of $Al_2O_3$: 8, 16, 24, 32, 80, 160; total number of ALD cycle of ZnO: 5, 10, 15, 20, 40, 80, and 160). The SH and TH signals from the MI samples (background signal) are attributed to the surface and bulk nonlinearities [16]. A solution containing Au nanoparticles (AuNPs) with 30 nm radius was then placed on the MI sample and distributed evenly over the surface by spin coating to produce MIM samples. The samples were dried using $N_2$ gas. The experimental procedure to measure SHG and THG from MIM samples follows the same recipe to that described above for MI samples. .

4. **RESULTS**

The results are reported for two thin film materials ($Al_2O_3$ and ZnO). Extensive experiments were performed using ALD deposition of surface films on an Au substrate to characterized and probe the samples. The SH signals for all our experiments follow the analogous behavior in Figure 3. The film thickness grows nearly linearly with the number of ALD cycles. Although, the linearity is accurate for larger number of ALD cycles, the linearity is estimated for smaller cycle numbers using the linear fit. For smaller film thicknesses the measured, surface film roughness (~1 nm) is large due to localized nucleation sites (See Supplement for ZnO data). For all sets of the same film material the SH signal shows an initial drop as the bare Au surface is covered with material followed by a plateau region when the surface is fully covered and the film becomes thicker. Reference [16] reported that the deposition of $Al_2O_3$ films progresses by a nucleation and growth mechanism; for the $AL_2O_3$ film complete surface coverage was achieved at around a film thickness of 4 nm corresponding to 32 ALD cycles. The ZnO films also show nucleation and growth, but the surface becomes completely covered by 2.4 nm or 10 ALD cycles. The rough surfaces means that the AuNPs may lie closer to the Au film for larger ALD cycles than measured by the linear fit for the film thickness.

$Al_2O_3$ films have no bulk second-order nonlinear contribution; this is contrasted with ZnO films, which has a nonzero one. The steep increase in the SH signal in ZnO in Figure 3(b) for the 30 nm thick film is attributed to the intrinsic ZnO nonlinearity. Its bulk nonlinear coefficients are widely discussed in the literature [25, 26]. The bulk, second-order susceptibility in ZnO estimated as $\chi^{(2)}(2\omega) = 1\ (pm/V)$ [27]. With this estimated value the bulk, second-order nonlinearity dominates the second-harmonic signals for thicker films. The SH efficiencies for the MIM data in Figure 3 show the same behavior as the MI samples; namely, a plateau is observed for all film thicknesses after the coverage is complete and there is no enhanced SH signal.

The TH efficiencies in Figure 4 reveal a marked difference between the MI and MIM samples. The bulk third-order nonlinearity of $Al_2O_3$ is small with a reported value $\chi^{(3)}(3\omega) = 23\ (pm^2/V^2)$ [27]. The bulk nonlinearities in

ZnO are large with $\chi^{(3)}(3\omega) = 1.85 \times 10^4 (pm^2/V^2)$. The MIM samples display a peak in the TH efficiency, which is attributed to the local field enhancement in the gap between the AuNPs and the Au film. The rise of the TH efficiency for the 30 nm thick ZnO film is due to its bulk nonlinear coefficient.

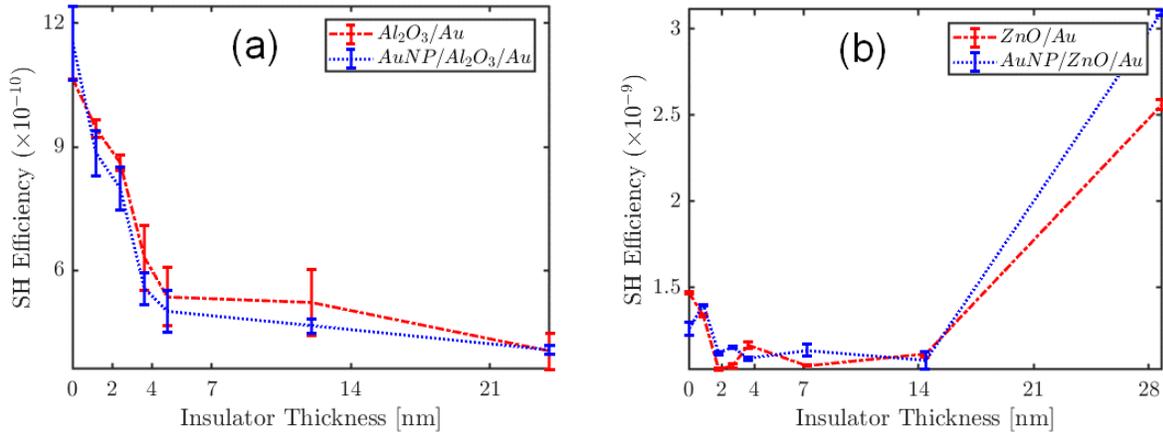

Fig. 3. SH efficiencies for (a) Al$_2$O$_3$ and (b) ZnO films for both MI and MIM samples. For Al$_2$O$_3$ the data points correspond to 8, 16, 24, 32, 80, 160 ALD cycles. For ZnO the data points correspond to samples with ideal surface coverage 5, 10, 15, 20, 40, 80, 160 ALD cycles.

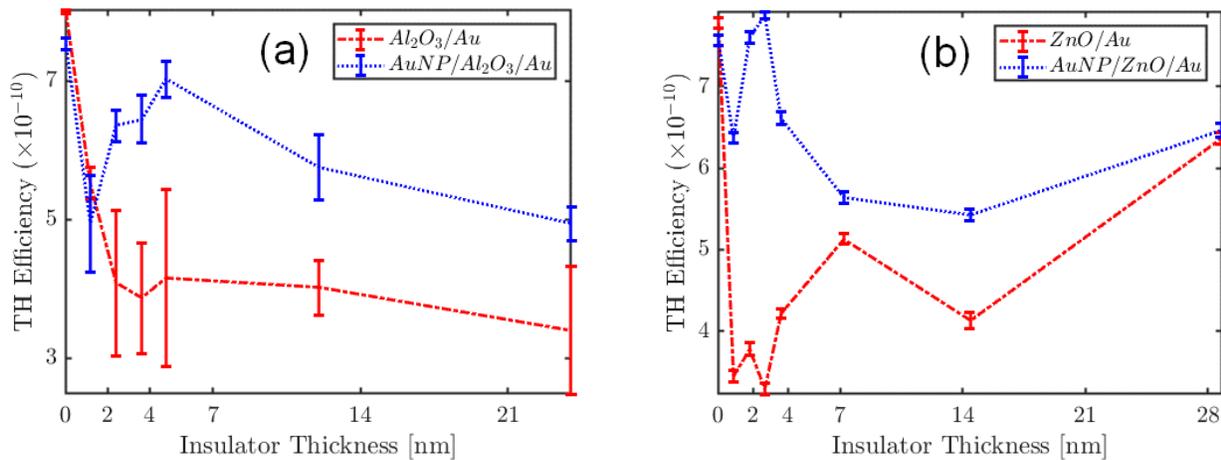

Fig. 4. TH efficiencies for (a) Al$_2$O$_3$ and (b) ZnO-Au (MI) for both MI and MIM samples. See Figure 3 for a discussion of the data points.

For Al$_2$O$_3$ films the TH efficiencies in Figure 4(a), a clear peak is shown at 5 nm (32 ALD cycles) for the MIM samples. This is the thickness at which the Al$_2$O$_3$ nuclei coalesce into a continuous film. The TH quenching continued down to 1 nm (8 ALD cycles). The bulk third-order nonlinearity contribution for Al$_2$O$_3$ was insignificant for all ALD cycles. In ZnO the peak is much sharper, which is consistent with the earlier coalescence of the nuclei in ZnO. The deposition proceeds by nucleation and growth, and we estimate (see Supplement), that the Au surface is fully covered with a ZnO film at 2.7 nm (10 ALD cycles) [28]. The simulation results discussed below infer that the peak enhancement factor is around a film thickness of 0.5 nm. The AuNPs are found in the lower positions on the surface, which means that on the average they are closer to the Au substrate than the dielectric film thickness measurement implies.

The SH signal from the gold film is mainly attributed to surface dipoles. The contribution from bound electrons is negligible and the free electrons have a quadrupole-like SH current density source in bulk Au. Denoting the pump (fundamental frequency) electric field as $\mathbf{E_1}$ the volume SH current density is

$$J_2^V = -2i\omega\varepsilon_0 b \frac{\beta}{2}\left[(a-1)(\chi_f \mathbf{E}_1 \cdot \nabla)\mathbf{E}_1 + \frac{\chi_f}{2}\nabla(\mathbf{E}_1 \cdot \mathbf{E}_1)\right]. \qquad (2)$$

The dipole-like SH surface current source ($J_2^S$) at the Au-insulator interface with tangential (x) and normal (y) components are,

$$\hat{\mathbf{x}} \cdot J_2^S = -2i\omega A\varepsilon_0 b \frac{\beta}{2}\chi_f(a+1)E_x E_y^{Au}, \qquad (3)$$

$$\hat{\mathbf{y}} \cdot J_2^S = 2i\omega B\varepsilon_0 b \frac{\beta}{2} E_y^{Au} E_y^{Au} \frac{\chi_f}{2\varepsilon_B}[\varepsilon_p + \varepsilon_B(2a+1)]. \qquad (4)$$

$E_y^{Au}$ is normal electric field component in Au, $\varepsilon_p$ is total dielectric constant of metal, due to both free and bound electrons (values for Au from Palik et al., [29]), $\varepsilon_B$ is dielectric constant of adjacent medium, and $\omega$ is angular frequency of pump signal. $\chi_f$ is the Drude model susceptibility, $a, \mathbf{b}$ and $\beta$ are related to the Drude model parameters, and $A$ and $B$ are Rudnick and Stern parameters. The parameter values used in our simulations are found in Reference [16]. The source for the TH signal in our simulations was the result in incorporating a bulk third-order coefficient. The results for SH and TH conversion efficiencies are plotted in Fig. 5. The simulations demonstrate the strong effect that the electron spill out has on the conversion efficiency. For a range of thicknesses of the insulator film the conversion efficiency is constant. The SH signal experiences a drop due to shielding of the normal component of the SH current density, $\hat{\mathbf{y}} \cdot J_2^S$, as indicated in the figure. The additional drop is due to the electr4on spill out. The TH efficiency drop is affected only by the electron spill out region and density. The electron density in the spill out region is 10% of the bulk free electron density in Au.

For the MIM data in Figure 3, the SH signal due to local field enhancement was not observed due to the dilute surface coverage of the AuNPs; the average separation between gold nanoparticles was estimated as 250 nm based on SEM measurements, meaning thousands of particles were captured within the spot size of the laser for a surface coverage < 2%. The initial drop of the SH signal at low coverage is due the screening of the surface dipoles by electrons that spill out into the insulator. The electrons form Metal-Induced Gap States (MIGS) that we reported earlier and contribute to the dipole screening [6, 16]. As the ZnO film approaches bulk thickness (~30nm) the SH efficiency grows, which is a consequence of its second-order bulk nonlinearity. The second-order nonlinearity due to the deposited ZnO film is sufficient to account for the observed increase at 30 nm (160 ALD cycles) [2, 6] (See Supplement)

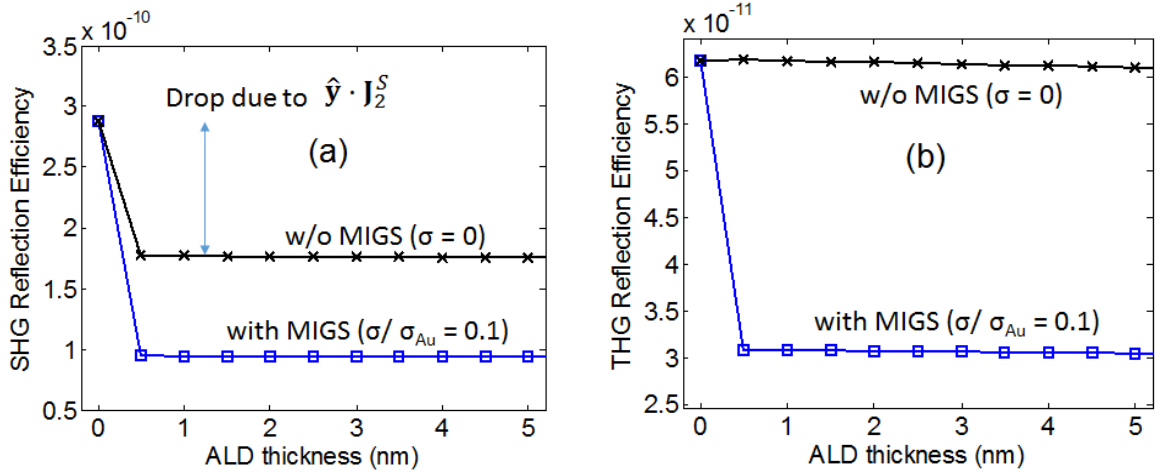

Fig. 5. (a) SH and (b) TH efficiencies of Al$_2$O$_3$-Au (MI) and Au-Al$_2$O$_3$-Au (MIM) samples. Data points correspond to 8, 16, 24, 32, 80, 160 ALD cycles.

The TH efficiency of the MI samples shows a precipitous drop in the signal at low coverage with large fluctuations in the TH efficiency where the SH signal is relatively flat. The increase in the TH efficiency for thicker ZnO films is due to the large third-order nonlinearity of ZnO [25, 26].

The MIM samples in Figure 4 show an enhanced TH efficiency above those for the MI samples. We observed the TH peak efficiency for the MIM sample at 2.7 nm (15 ALD cycles). The TH efficiency has a local maximum for the exposed Au surface; as ZnO is deposited on the surface MIGS create an electron screening effect that initially reduces the TH signal. For the MIM samples a minimum in the TH signal was observed at 0.9 nm (5 ALD cycles). The TH efficiency grows sharply for thicker films.

To understand the observed THG behavior in MIM samples extensive simulations using gold nano-spheres were performed using finite element methods. The local field enhancement results for four cases are plotted in Figure 6. The simulations included both quantum and nonlocal contributions (see Supplement). The classical curve omits both quantum and nonlocal effects. For the curve labeled "quantum" the simulations only included the quantum conductance; it has a peak around 0.3 nm ZnO layer thickness with a peak value reduced in comparison with the classical curve. For the curve labeled "nonlocal" our simulations include only the nonlocal effects; the peak enhancement factor occurred when the thickness of the ZnO film was about 0.5nm. This indicates that nonlocal contribution dominate over the quantum contribution down to about 0.2 nm.

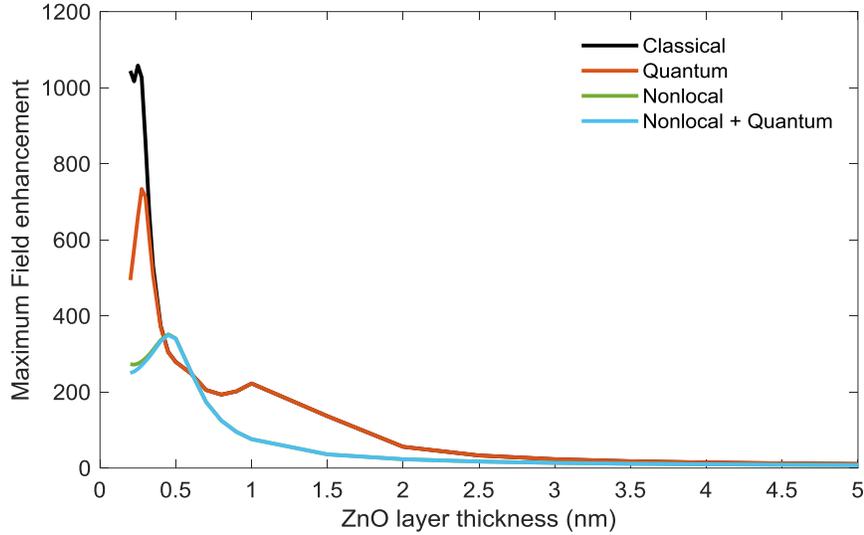

Fig. 6. Comparison between four different simulations. The Classical results omit nonlocality and quantum effects; the Quantum effects incorporate electron tunneling and the Nonlocal calculation incorporates spatial dispersion effects. The Nonlocal + Quantum curve includes both effects.

Comparison between the TH MIM experimental results and the enhancement factor from the simulations shows a discrepancy in the film thickness. As discussed above, we resolve this discrepancy by noting that the nucleation and growth of the thin film roughens the film surface and nanoparticles settle in the valleys between the ZnO and $Al_2O_3$ islands. The nanoparticles have a range of gap thicknesses and on average the gap is much smaller than the 2.7 nm and 5 nm thicknesses derived from the linear fit. Using the maximum as a determination of the average AuNP/Au film separation of 0.5 nm from our simulation, the we infer a much closer approach of the nanoparticles This is consistent with our findings that dielectric films have a nucleation and growth coalescence.

## 5. CONCLUSION

For both Au-$Al_2O_3$-Au and Au-ZnO-Au samples, the SH efficiencies in Figure 3 show a signal reduction and plateau. The TH efficiencies for MI samples show the same signal reduction and plateau phenomenon and are in accordance with our simulations. The results corresponds to our expectations based on simulations of the electron spill out into the insulator, which is the MIGS effect.

The gradual drop observed for SH efficiency in $Al_2O_3$ MI samples with increasing number of ALD cycles was reported earlier [16], where the plateau value for the SH signal was attributed to electron spill out from metal into the insulator and promotes the formation of MIGS. The ALD deposited ZnO thin film is more conformal and hence, the full coverage is observed by SH signal for 10 ALD cycle number.

While electron quantum tunneling is observed in MIM structures called rectennas (rectifier + antennas) and capacitors we determined by simulations that nonlocal effects and not quantum tunneling are the dominant mechanism limiting the local field enhancement in our experiments. The TH efficiencies for both insulator materials show a characteristic signal peak and subsequent quenching for the thinnest insulator films. The SH and TH generated from MI and MIM interfaces can be applied to reveal information about E-field enhancement and the spill out electron density at the metal/insulator interface.

**Funding**. Full support from U.S. Army RDECOM (W911NF-15-1-0178) is acknowledged. Research of DdC was sponsored by the RDECOM-Atlantic, US Army Research Office, and Office of Naval Research Global and was partly accomplished under Grant Number W911NF-18-1-0424.

**Acknowledgment**. MMRH gratefully acknowledges '2017 Graduate Student Summer Fellowship' support from the University of Dayton. The characterization facilities of the Institute of Materials Science & Engineering at Washington University in St. Louis are gratefully acknowledged.

See Supplement for supporting content.

**REFERENCES**
1. F. J. Garcia de Abajo, "Nonlocal effects in the plasmons of strongly interacting nanoparticles, dimers, and waveguides," J. Phys. Chem. B **112**, 17983 (2008).
2. S. Raza, S. I. Bozhevolnyi, M. Wubs, and N. A. Mortensen, "Nonlocal optical response in metallic nanostructures," J. Phys. Cond. Matt. **27**, 183204 (2015).
3. J. M. McMahon, S. K. Gray, and G. C. Schatz, "Optical properties of nanowire dimers with a spatially nonlocal dielectric function," Nano Lett. **10**, 3473 (2010).
4. C. Ciraci, R. T. Hill, J. J. Mock, Y. Urzhumov, A. I. Fernandez-Dominguez, S. A. Maier, J. B. Pendry, A. Chilkoti, and D. R. Smith, "Probing the ultimate limits of plasmonic enhancement," Science **337**, 1072 (2012).
5. Y. Luo, A. I. Fernandez-Dominguez, A. Wiener, S. A. Maier, and J. B. Pendry, "Surface plasmons and nonlocality: a simple model," Phys. Rev. Lett. **111**, 93901 (2013).
6. M. Scalora, D. de Ceglia, M. A. Vincenti, and J. W. Haus, "Nonlocal and quantum tunneling contributions to harmonic generation in nanostructures: electron cloud screening effects," Phys. Rev. A **90**, 013831 (2014).
7. N. A. Mortensen, S. Raza, M. Wubs, T. Søndergaard, and S. I. Bozhevolnyi, "A generalized non-local optical response theory for plasmonic nanostructures," Nature Comm. **5**, 3809 (2014).
8. J. Zuloaga, E. Prodan, and P. Nordlander, "Quantum description of the plasmon resonances of a nanoparticle dimer," Nano Lett. **9,** 887 (2009).

9. J. Zuloaga, E. Prodan and P. Nordlander, "Quantum plasmonics: optical properties and tunability of metallic nanorods," ACS Nano **4**, 5269 (2010).

10. D.C. Marinica, A.K. Kazansky, P. Nordlander, J. Aizpurua, and A.G. Borisov, "Quantum plasmonics: nonlinear effects in the field enhancement of a plasmonic nanoparticle dimer," Nano Lett. **12**, 1333 (2012).

11. R. Esteban, A.G. Borisov, P. Nordlander, and J. Aizpurua, "Bridging quantum and classical plasmonics with a quantum-corrected model," Nature Comm. **3**, 825 (2012).

12. J. W. Haus, L. Li, N. C. Katte, C. Deng, M. Scalora, D. de Ceglia, and M. A. Vincenti, "Nanowire metal-insulator-metal plasmonic devices," Prathan Buranasiri, ed. Proceedings of SPIE **8883**, 888303 (2013).

13. T. V. Teperik, P. Nordlander, J. Aizpurua, and A. G. Borisov, "Quantum effects and nonlocality in strongly coupled plasmonic nanowire dimers," Opt. Express **21**, 27306 (2013).

14. J. W. Haus, D. de Ceglia, M. A. Vincenti, and M. Scalora, "Quantum conductivity for metal-insulator-metal nanostructures," J. Opt. Soc. Am. B **31**, 259 (2014).

15. J. W. Haus, D. de Ceglia, M. A. Vincenti, and M. Scalora, "Nonlinear quantum tunneling effects in nano-plasmonic environments," J. Opt. Soc. Am. B **31**, A13-A19 (2014).

16. Z. Gao, M. M. R. Hussain, D. de Ceglia, M. A. Vincenti, A. Sarangan, I. Agha, M. Scalora, J. W. Haus, and P. Banerjee, "Unraveling delocalized electrons in metal induced gap states from second harmonics," Appl. Phys. Lett. **111**, 161601 (2017).

17. G. Hajisalem, M. S. Nezami, and R. Gordon, "Probing the quantum tunneling limit of plasmonic enhancement by third harmonic generation," Nano Lett. **14**, 6651-6654 (2014).

18. W. Zhu, and K. B. Crozier, "Quantum mechanical limit to plasmonic enhancement as observed by surface-enhanced Raman scattering," Nature Comm. **5**, 5228 (2014).

19. W. Zhu, R. Esteban, A. G. Borisov, J. J. Baumberg, P. Nordlander, H. J. Lezec, J. Aizpurua, and K. B. Crozier, "Quantum mechanical effects in plasmonic structures with subnanometre gaps," Nature Comm. **7**, 11495 (2016).

20. S. Grover, and G. Moddel, "Engineering the current–voltage characteristics of metal–insulator–metal diodes using double-insulator tunnel barriers," Solid-State Electronics **67**, 94 (2012).

21. M. Kauranen, and A. V. Zayats, "Nonlinear plasmonics," Nature Phot. **6**, 737 (2012).

22. M. Scalora, M. A. Vincenti, D. de Ceglia, C. M. Cojocaru, M. Grande, and J. W. Haus, "Nonlinear Duffing oscillator model for third harmonic generation," J. Opt. Soc. Am. B **32**, 2129 (2015).

23. P. E. Powers and J. W. Haus, *Fundamentals of Nonlinear Optics* 2nd Ed., (CRC Press, 2017).

24. S. Kawata, Y. Inouye, and P. Verma. "Plasmonics for near-field nano-imaging and superlensing," Nature Phot. **3**, 388 (2009).

25. M. C. Larciprete, D. Haertle, A. Belardini, M. Bertolotti, F. Sarto, and P. Günter. "Characterization of second and third order optical nonlinearities of ZnO sputtered films," Appl. Phys. B: Lasers and Optics **82**, 431 (2006). Correction: $\chi^{(3)}_{3333} = 1.85 \times 10^4 \, pm^2/V^2$.

26. M. C. Larciprete, and M. Centini. "Second harmonic generation from ZnO films and nanostructures," Appl. Phys. Rev. **2**, 031302 (2015).
27. J. Britt Lassiter, X. Chen, X. Liu, C. Ciracì, T. B. Hoang, S. Larouche, S.-H. Oh, M. H. Mikkelsen, and D. R. Smith, "Third-harmonic generation enhancement by film-coupled plasmonic stripe resonators," ACS Phot. **1**, 1212 (2014).
28. Z. Gao, F. Wu, Y. Myung, R. Fei, R. Kanjolia, L. Yang, and P. Banerjee, "Standing and sitting adlayers in atomic layer deposition of ZnO." J. Vac. Sci. & Tech. A: Vacuum, Surfaces and Films **34**, 01A143 (2016).
29. E. D. Palik, and G. Ghosh, *Handbook of optical constants of solids*. (Academic press, 1998).

# Harmonic Generation in Metal-Insulator and Metal-Insulator-Metal Nanostructures: supplementary material

MALLIK M. R. HUSSAIN[1], IMAD AGHA[1,*], ZHENGNING GAO[2], DOMENICO DE CEGLIA[3], MARIA A. VINCENTI[4], ANDREW SARANGAN[1], MICHAEL SCALORA[5], PARAG BANERJEE[2,6], JOSEPH W. HAUS[1]

[1]Department of Electro-Optics and Photonics, University of Dayton, OH - 45469, USA

[2]Department of Materials Science & Engineering, University of Central Florida, Orlando, FL - 32816, USA

[3]Department of Information Engineering, University of Padova, Padova, Italy,

[4]Department of Information Engineering, University of Brescia, Brescia, Italy

[5]US Army AMRDEC, Charles M. Bowden Research Laboratory, Redstone Arsenal, AL - 35898, USA

[6]Department of Mechanical Engineering & Materials Science, Washington University, St. Louis, MO - 63130, USA

*Corresponding authors: jhaus1@udayton.edu, iagha1@udayton.edu

We present additional, detailed information about our simulations and experiments in this supplement.

## 1. PHOTON-ASSISTED-TUNNELING' THEORY

The 'photon-assisted-tunneling' (PAT) theory [1-3] utilizes the mathematical formulation of many-body physics to derive the relationship between the microscopic 'probability current density', $J_P$, of electronic wave function (in the presence of single or multiple photons) and the macroscopic DC and AC 'current density', $J(t) = \sum_{m=0,1,2,...} Re\{(J_{m\omega}/2)\, e^{j \cdot m\omega t}\}$. Here, $\omega$ is the incident photon frequency, $t$ is time, $J_{m\omega}$ is the amplitude of rectified ($m=0$) current density or AC current densities of fundamental ($m=1$) or higher harmonic frequencies ($m=2,3,...$ etc.) of $\omega$. For the case of metal-insulator-metal (MIM) tunnel junctions [4-8], the equation relating the microscopic and macroscopic counterparts of current densities can be written as follows:

$$\boldsymbol{J}_{m\omega} = \sum_{n=-\infty}^{+\infty} B_n(\alpha)[B_{n+m}(\alpha) + B_{n-m}(\alpha)] \cdot \boldsymbol{J}_P(n\hbar\omega + qV_{bias}). \qquad (S1)$$

Here, $\boldsymbol{B_n}$ is the $n^{th}$ order Bessel function, $V_{bias}$ is the DC bias voltage across MIM interface, $q$ is the electronic charge. Also,

$$\alpha = \frac{Electron\ energy}{Photon\ energy} = \frac{qwE_\omega}{\hbar\omega}, \qquad (S2)$$

where, $\alpha$ is the ratio of electron energy difference (between two metal ends of MIM interface) and photon energy, $w$ is the insulator thickness, $E_\omega$ is E-field amplitude of the incident photon frequency inside the insulator section of the MIM device. Classically, $E_\omega$ inside the insulator can increase unboundedly (up to infinity) as the thickness of the insulator gets smaller. But, quantum mechanically, the E-field enhancement is limited by the on-set of tunneling current. The AC part of the tunneling current (sloshing of electrons from one metal to another through the insulator) couples to the electromagnetic radiation that can be detected in the far field. Such radiation from MIM interface adds to the background radiation that originates from the surface and bulk nonlinearities of metal-insulator (MI) interfaces. The goal of this paper is to detect the electromagnetic radiation of second and third harmonic (SH and TH) frequencies for $V_{bias} = 0$ across the MIM interfaces which is in excess of the background contribution from MI interface.

For SH and TH generation (i.e. $m = 2, 3$), $n = 0, \pm 1, \pm 2, \pm 3$ were considered. In such cases, Bessel functions can be expanded, and Equations S1 can be rewritten as,

$$J_{2\omega} = 0, \tag{S3}$$

$$J_{3\omega} = \frac{\alpha^3}{8}\left[J_P(\omega) - J_P(2\omega) + \frac{1}{3}J_P(3\omega)\right]. \tag{S4}$$

It is apparent from Equation S3 that tunneling at the MIM interface has no contribution in SH generation. From Equation S4, we infer that, there are two competing variables, $\alpha$ and $J_P$, both of which depend on $\omega$. For a large $w$, $J_P \to 0$ and photons with third harmonic (TH) frequency does not change from MI to MIM. As $\omega$ is gradually made smaller (thinner insulator), the $J_{3\omega}$ peaks ($E_{3\omega}$-field enhancement threshold) for certain $w$, thus TH signal from MIM junction peaks. Finally, for $w \to 0$, $\alpha \to 0$ and TH signal from MIM again drops to MI level (field quenching).

Our calculations of this for insulators ZnO and $Al_2O_3$ yield maximum values in the range $\chi^{(3)}(3\omega) \sim 1\ (pm/V)^2$. This value is much smaller than the intrinsic nonlinear susceptibility of ZnO, but is comparable to the bulk coefficient value in $Al_2O_3$. We determined that the main effect of quantum tunneling in our experiments is the linear conductance coefficient. The AC tunneling current couples directly to the fundamental frequency field.

This experiment was done for several MIM samples (Au-$Al_2O_3$-Au and Au-ZnO-Au) with varying insulator thicknesses. The predictions of PAT (i.e. no change in SH efficiency and enhancement and quenching for TH efficiency due to tunneling) were observed in our experiment.

The quantum conductivity was calculated for both ZnO and $Al_2O_3$ and inserted into our COMSOL simulations. The figures below are the calculated quantum susceptibilities using the relation $\chi^{(1)} = i\sigma/\omega\varepsilon_0$. It is purely imaginary in our calculations. In Fig. S1 the susceptibility amplitudes are comparable and have essentially the same effect on the incident electromagnetic field.

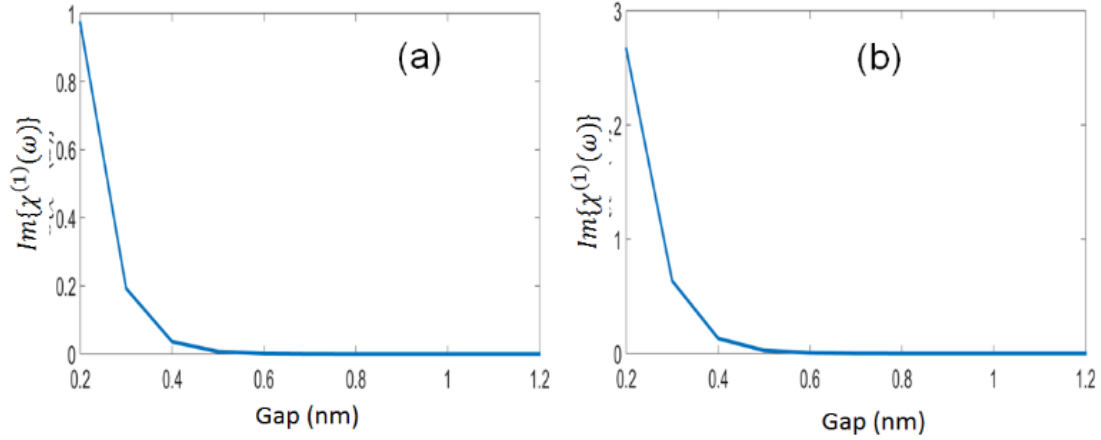

Fig. S1. The linear quantum susceptibilities for $Al_2O_3$ and ZnO, respectively. The wavelength is 810 nm.

## 2. EXPERIMENTAL SETUP

Our experimental setup is illustrated in Fig. S2. All the components of our system are discussed in this section. Three lasers were collinearly aligned to pass along the same path from the sample to the photodetector. The HeNe laser (632.8 $nm$) was used to fix the optic axis (OA). A blue laser (405 $nm$) was used to verify prism alignment for the SH signal. The mode-locked Ti:Sapphire laser (810 $nm$) was used as the input fundamental frequency laser (or pump laser). The laser had a pulse width of 80 $fs$ and pulse repetition rate of 80 $MHz$ and output power of 400 mW. The L1 lens (Thorlabs A260TM-B, f = 15.31 $mm$ @ 810 $nm$, AR coating: 650-1050 $nm$) focused the pump beam. The beam waist (at the focal position) was measured to be 25.49 $\mu m$ in horizontal direction and 23.45 $\mu m$ in vertical direction. At the focal position the FF laser, intensity was calculated to be 6.1574 $GW/cm^2$. A half wave plate (HWP) rotated the FF polarization; p-polarization with respect to the sample surface was used to collect the experimental data. Samples were placed on a stage with required degrees of freedom to confirm that they were at the focal position of L1 (at 810 $nm$). Two achromatic doublet lenses (with AR coating: 240-410 $nm$), i.e. L2 (Thorlabs ACA254-150-UV, f = 150 $mm$) and L3 (Thorlabs ACA254-100-UV, f = 100 $mm$) was used to collimate and refocus the signal reflected from the sample surface. The focal lengths of L1, L2 were so chosen that the effective numerical aperture (NA) of the incident beam matches that of the reflected beam (i.e. $NA_{eff,i} = NA_{eff,r} = 0.0653$) and thus, the collection efficiency of the output arm is reasonably high. The choice of focal length of L3 (compared to L2) ensured a demagnifying system that increases the system's tolerance to sample tip-tilt and defocus. Two UV grade synthetic fused silica (UVGSFS) prisms were used to spatially separate the SH and TH from the much intense FF. Both prisms were placed on rotating stages and their angle of deviation for different incident angles at different wavelengths were utilized to select pairs of incident angles for SH and TH light. Filters (Thorlabs: FESH0450 for SH and Semrock: FF01-276 for TH) were used to further clean the signal. The filters were placed in the collimated beam path to reduce ghost images and increase filtering efficiency. A TE-cooled UV-enhanced Si detector (EO Systems: UVS-025-TE2-H) was placed at the focal length of L3 to collect the signal. The detector's (PD) output was, then, measured through a lock-in amplifier (LA), SR830. The detector was cooled to $-40^\circ C$ and a low gain configuration was used. The chopping frequency was set around 311 $Hz$. The chopper had a duty cycle of 50%. The output from PD was measures in units of voltage. Data were collected at the sample rate of 20Hz for 10-30 seconds to measure a time-average. For each measurement, the FF was reduced using calibrated ND filters and SH and TH signals were recorded. The power law dependences of signal intensity (quadratic for SH and cubic for TH) were used to verify the generation and detection of SH and TH.

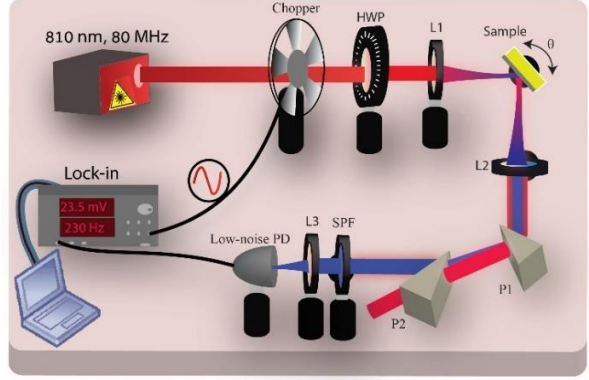

Fig. S2. Schematic of setup for measuring second and third harmonic signal generated on samples.

The transmittances of the output path of the system (with all the components i.e. L2, L3, P1, P2 and SPF in the path) were measured to be $T_{out,SH} = 0.46$ and $T_{out,TH} = 0.16$. These transmittances match the expected system transmittance calculated from manufacturer's transmittance data for each component in the system. The responsivity, $R_{PD,SH}$, of the detector for SH frequency was measured (and cross-checked with the manufacturer's data) to be $0.185$ $A/W$. The responsivity at TH, $R_{PD,TH}$, frequency was collected from the manufacture's data to be $0.15$ $A/W$. The transimpedance gain, $G$, of the amplifier inside the detector was $10^8$ $\Omega$. The lock-in amplifier measures RMS voltage from PD and it only phase locks to the fundamental frequency of modulation. For square-wave modulation, the lock-in amplifier constant, $C_{sq} = (\pi\sqrt{2})/4$. Therefore, the effective responsivity [$V/W$] of the output path is,

$$R_{eff} = \frac{G \cdot T_{out}}{C_{sq}} R_{PD}. \tag{S5}$$

For SH, the effective responsivity, $R_{eff,SH} = 7.6617 \cdot 10^6$ $V/W$. For TH, the effective responsivity, $R_{eff,TH} = 2.2405 \cdot 10^6$ $V/W$. For a single measurement, the voltage reads from lock-in amplifier, $V_{LA}$ are averaged over the total number of samples. The SH/TH power generated on the sample can be calculated as,

$$P_{SH/TH} = \frac{1}{R_{eff,SH/TH}} \langle V_{LA} \rangle. \tag{S6}$$

This $P_{SH/TH}$ was calculated for the whole set of samples treated with different numbers of ALD cycles. Finally, to determine the efficiencies of SH/TH generation following equation was used.

$$\eta_{SH/TH} = \frac{P_{SH/TH}}{P_{FF}}. \tag{S7}$$

## 3. CALCULATION OF $\chi^{(2)}$ FOR THE ZnO-Au STRUCTURE

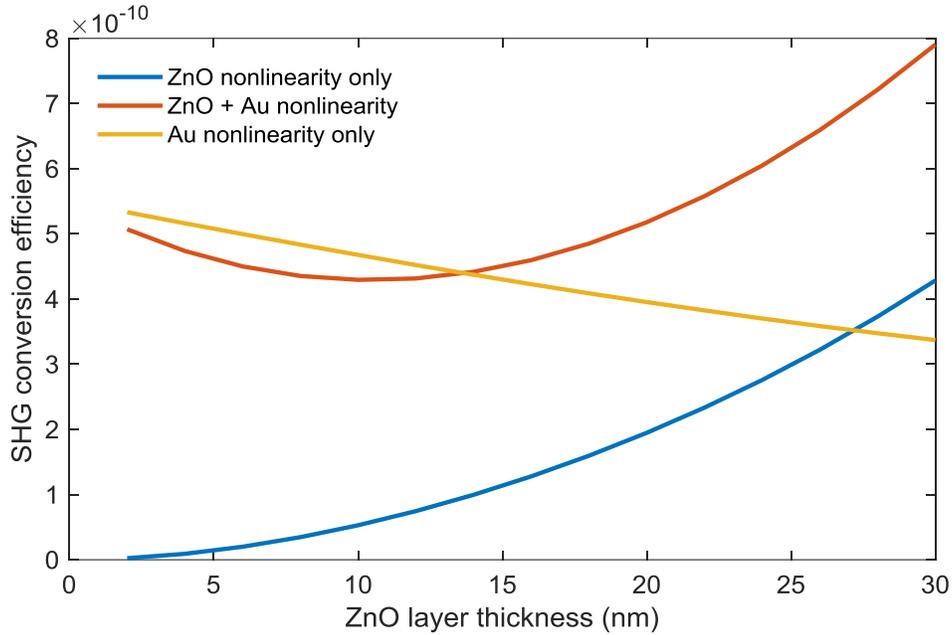

Fig. S3. SHG conversion efficiency from an MI structure i.e., ZnO film on a planar and opaque Au film.

SHG in both the MI and MIM structures originates mainly from the ZnO bulk ($Al_2O_3$ displays a vanishing $\chi^{(2)}$ in the bulk) and from Au surfaces. Here we determine the role of these two nonlinearities for different ZnO-film thicknesses in the samples investigated in our SHG experiments.

In Fig. S3, the SHG is calculated under three different circumstances: (i) assuming that only the ZnO nonlinearity is active; (ii) assuming that both the Au-surface and the ZnO nonlinearities are active; (iii) and assuming that only the Au-surface nonlinearity is active. The Au-surface nonlinearity is obviously important for very small values of the ZnO layer thickness ($<$ 10 nm), where the SH intensity slightly decreases as a function of the film thickness. On the other hand, an opposite trend is observed for ZnO film thicknesses larger than 10 nm, where the SH intensity grows with the film thickness. The latter behavior indicates a stronger importance of the ZnO bulk nonlinearity for large ZnO thicknesses, as confirmed by the very similar slope of the red [case (ii)] and blue curves [case (i)] in this regime.

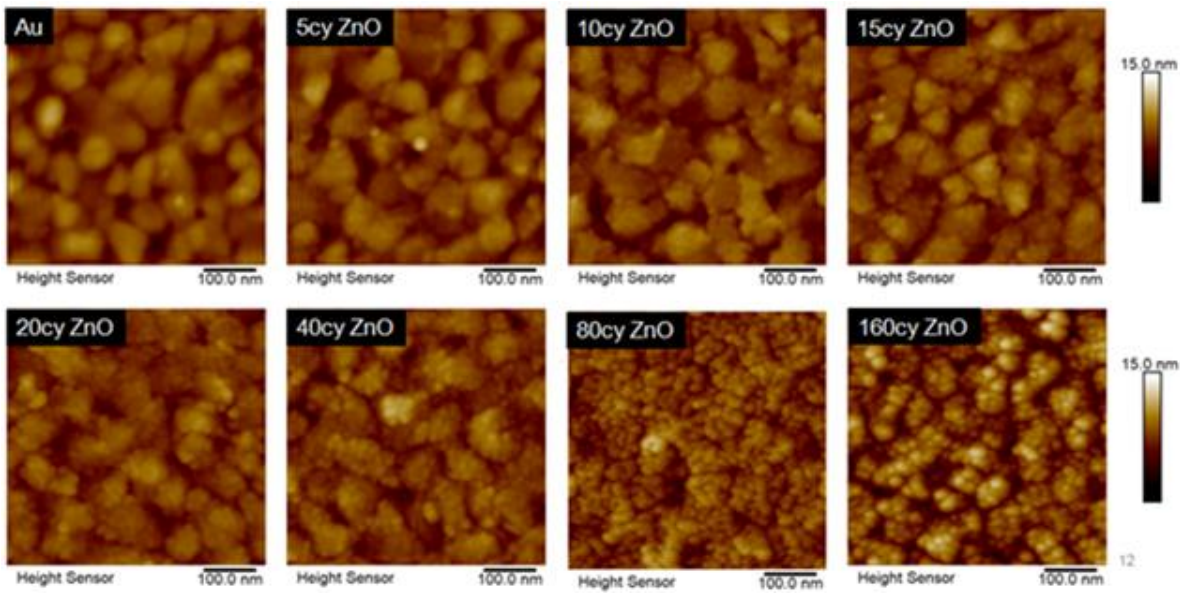
Fig. S4. AFM data of ZnO deposition on Au as a function of number of ALD cycles.

The efficiency is calculated as a function of the ZnO film thickness under TM-polarized, plane-wave illumination at 68° oblique incidence with pump peak irradiance equal to 6.15 GW/cm$^2$. The bulk ZnO nonlinearity is modeled with a quadratic susceptibility $\chi^{(2)} = 1$ pm/V. It is important to mention that, although the ZnO nonlinearity provides good a qualitative model of SHG for ZnO-film thicknesses larger than 10 nm, the Au-surface nonlinearity is equally important in this regime in order to provide a more accurate prediction of SHG conversion efficiency. In fact, the results in Fig. S3, in particular the model that includes both the quadratic nonlinearities [case (ii)] are in good agreement with the experimental data of SHG reported in Fig. 3(b) for the ZnO MI structure.

**ATOMIC FORCE MICROSCOPY (AFM) DATA**

AFM images of our samples are displayed in Fig. S4. In the images crystalline ZnO grains are observed to grow conformally on Au after 10 cycles. This growth during crystallization is obvious for 40, 80 and 160 cycles. The 'pillbox' model that correlates root mean square (rms) roughness to the nuclei morphology is discussed in detail in our previous publication [9].

The rms roughness for the ZnO-Au sample as a function of ALD cycle number is extracted from the AFM data (Fig. S5). A peak roughness at 5 cycles indicates nuclei formation and impingement. Beyond that, increase in ALD cycles results in lower film roughness as nuclei grow and coalesce into a continuous film. After 20 cycles, film-crystallization contributes to rms roughness. Thus for ZnO on Au, quick nucleation and conformal films are deposited, indicating low barrier to nucleation (compared to $Al_2O_3$).

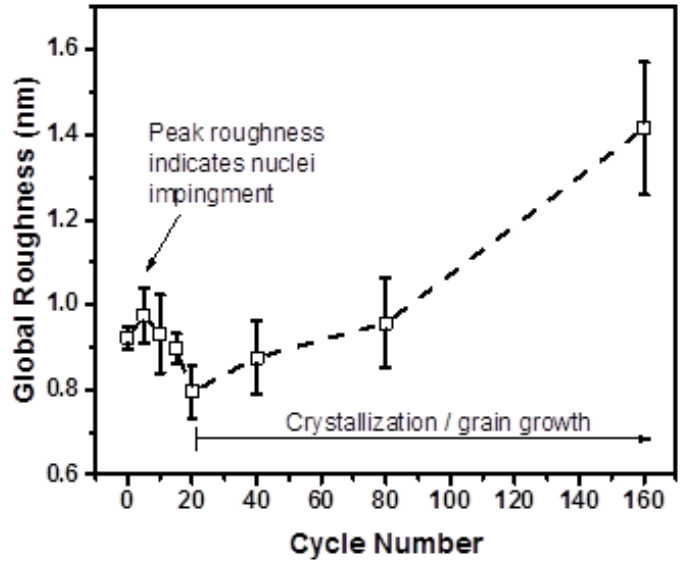

Fig. S5. Surface roughness after ALD deposition of ZnO. The results are derived from our AFM measurements.

## 4. CHARACTERIZATION MEASUREMENT OF ZnO

The thickness of the ALD grown films essentially follows a linear relationship, as shown in Fig. S6. ZnO grown on silicon is more conformal and the thickness grows at a lower rate. Spectroscopic ellipsometry is used to measure ZnO thickness as a function of number of ALD cycles on Au as well as Si with native $SiO_2$. The deposition rate (i.e., slope) of ZnO on Au is measured to be 0.18 nm/cycle whereas, the deposition rate of ZnO on Si is 0.16 nm/cycle.

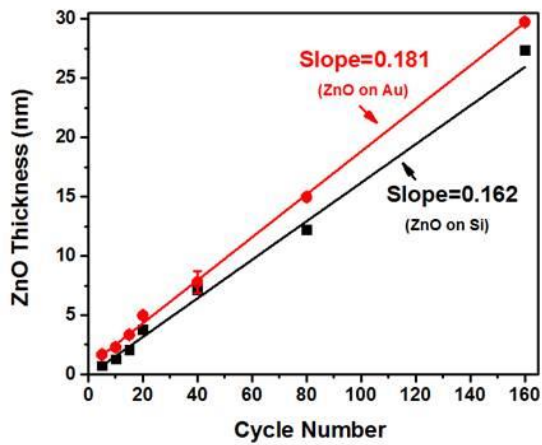

Fig. S6. Ellipsometric measurement of the ALD deposited ZnO film thicknesses. The ALD cycle number versus ZnO film thickness on gold (Au) and on silicon (Si).

## 5. XPS VALENCE BAND SPECTRA OF ZnO-Au AND Al2O3 –Au SAMPLES

Valence band, x-ray photoelectron spectra (XPS) is a highly sensitive, surface characterization technique that can indicate the relative fraction of surface coverage of a film over a substrate. This is because only exposed (or nearly exposed i.e., < 1 nm of film or less) surfaces contribute to the valence band spectra. Heterogeneous surfaces are responsible for producing spectral features with multiple peaks. We observe (Fig. S7) that for the ZnO–Au samples, the Zn 3s peak at ~10 eV emerges immediately after 5 ALD cycles [10]. The Au valence band features are completely quenched by 20 ALD cycles. This implies that a fully conformal film of ZnO covers the Au surface at 20 cycles and beyond.

Whereas, for the $Al_2O_3$–Au samples, the Au valence band features can be seen clearly at 32 cycles while completely removed only at 80 cycles. The Au valence band spectrum is replaced by Al 2s and O 2s peaks [11]. Thus for $Al_2O_3$–Au samples, complete surface coverage takes longer than 32 cycles. In our previous publication [9], we estimate, using AFM data, the number of $Al_2O_3$ ALD cycles for complete surface coverage to be 47.

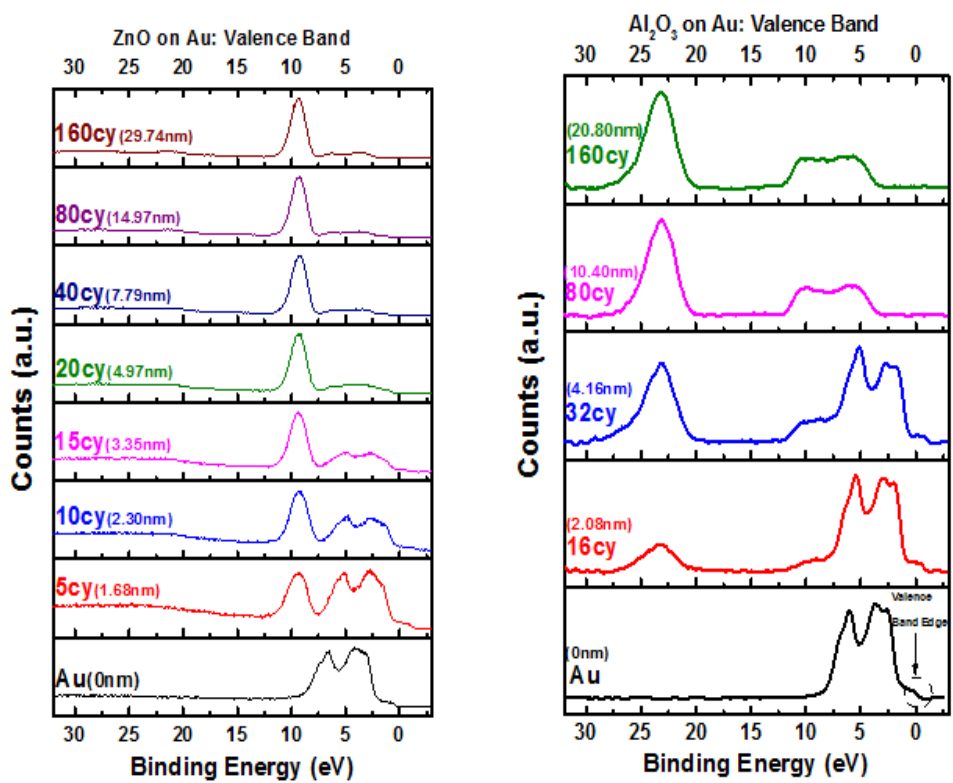

Fig. S7. X-ray photoelectron spectra for ALD deposited films ZnO and $Al_2O_3$ on u surfaces. The figure compares left, the valence band of Au upon ZnO deposition and right, the valence band of Au upon $Al_2O_3$ deposition. The vertically stacked spectra are for various ALD cycles.

6. **NANOPARTICLE DENSITY**

The SEM image in Fig. S8 highlights the gold nanoparticles (in green circles) sitting on the MI surface. We used circular Hough transformation to automatically identify nanoparticles on the SEM images. From these images we deduce the average separation between gold nanoparticles is 200 nm and 16.5 NPs/$\mu m^2$. The nanoparticles generally sit in the surface low regions.

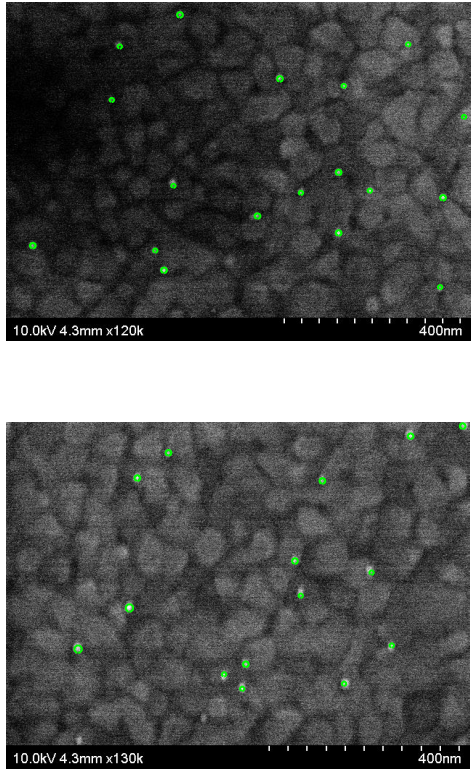

Fig. S8. Sample SEM images of gold nanoparticles (marked in green circles) distributed on a MI surface.

## 7. 3D Simulations

The field enhancement at the pump wavelength has been calculated in the MIM structure with the finite-element-method (COMSOL). We have investigated the effects of the quantum tunneling current through the insulator gap and the nonlocal response of the free electrons in the metal. Quantum tunneling has been introduced as a conductivity in the insulator gap as in [josab Joe paper], namely by altering the gap permittivity as follows, $\epsilon_{q,gap} = \epsilon_{c,gap} + \frac{i\sigma_{gap}}{\omega\epsilon_0}$, where $\epsilon_{c,gap}$ is the gap permittivity without quantum currents and $\sigma_{gap}$ is the wavelength- and gap-thickness-dependent conductivity associated with quantum tunneling. In the hydrodynamic theory of free electrons in metals, the nonlocality arises from spatial variations of the electron pressure $\nabla p$, and it leads to a spatially-dispersive optical response [12]. In other words, free-electrons currents are also susceptible to electric-field spatial derivatives. This effect is significant near metal surfaces, edges and corners, i.e., in regions characterized by strong charge accumulation and hence large values of field spatial derivatives. In COMSOL, the nonlocality is introduced by modifying the Ohm's law in the metallic domains as follows,

$$\beta^2 \nabla(\nabla \cdot \boldsymbol{J}_\omega) + \omega^2 \boldsymbol{J}_\omega + i\omega\gamma \boldsymbol{J}_\omega = i\omega\varepsilon_0 \omega_p^2 \boldsymbol{E}_\omega, \tag{S8}$$

in which $\boldsymbol{J}_\omega$ and $\boldsymbol{E}_\omega$ are the induced current and the electric field phasors in the metal, $\beta$ is the nonlocal coefficient, $\gamma$ is the collision frequency for electrons and accounts for damping, and $\omega_p$ is the bulk plasma frequency. For the kinetic pressure, we adopt the expression of a degenerate Fermi gas at T = 0 K, hence the electron pressure is polytropically related to the electron density, namely with $p = \frac{\hbar^2}{5m^*}(3\pi^2)^{2/3} n^{5/3}$. In this model, the nonlocal coefficient is $\beta =$

$\sqrt{\frac{3}{5}} v_F$, where $v_F \sim 10^6$ m/s is the electrons Fermi velocity in noble metals as gold and silver. A high mesh resolution is required to accurately account for quantum and nonlocal effects both in the insulator and in the metallic regions, especially when the gap is smaller than a few nanometers. For this reason, we exploit the axial symmetry of the structure and solve our problem of a gold sphere on a planar thin film structure in two dimensions instead of three, by using cylindrical coordinates and considering a r-z plane with azimuthal angle equal to zero. This requires an expansion of the obliquely incident plane wave (the angle of incidence is 68 °) source and the scattered fields in cylindrical harmonics, following the procedure outlined in [13]. Since the field is mostly confined near the r=0 axis, only few cylindrical harmonics are required in order to find an accurate solution near the gap region.

In Fig. S9 we report the maximum field enhancement in the gap region as a function of the gap thickness for the case of an Au nanosphere on top of an opaque gold layer separated by a ZnO film. The nanoparticle radius is 30 nm and a plane wave of 810 nm wavelength illuminates the structure at with an incident angle of 68 °. The maximum field enhancement is plotted under four different circumstances: (i) fully classical model (i.e., absence of quantum currents and nonlocality); (ii) presence of quantum tunneling in the gap; (iii) presence of nonlocality in gold; (iv) simultaneous presence of quantum and nonlocal effects. It is important to mention that the field distribution in the region is not uniform due to the strong dependence of the field enhancement with respect to the distance between the metallic surfaces facing the gap region: given the nanoparticle spherical shape, this distance obviously shows its minimum at r=0 and grows as a quadratic function of r. This means that, even in case (i), i.e., within the fully classical theory, there is a non-trivial dependence of the field-enhancement on the gap-region thickness, due to the presence of multiple plasmonic modes localized within the MIM structure. It is clear that quantum effects are only relevant below 0.3 nm for this structure, while nonlocal effects are significant even at gap sizes as large as 5 nm.

The distinction between case (iv), i.e., nonlocal and quantum effects simultaneously present, and the case (iii), i.e., only nonlocality present, is barely noticeable in our simulations. This means that, although both quantum and nonlocal effects tend to limit the field enhancement for very narrow gap sizes, the main quenching mechanism in our structure is by far the nonlocality. The nonlocality strength is due to the fact that the spatially-dispersive term, $\beta^2 \nabla(\nabla \cdot \boldsymbol{J}_\omega)$, term induces an effective blueshift of the plasma frequency on the order of $\beta |\mathbf{k}|$, where the wave vector k stems from the transformation $\nabla \rightarrow i\mathbf{k}$ into the k-space domain. As a consequence, resonances associated with the MIM plasmonic modes may be significantly blueshifted (tens of nanometers) and hence may induce large field-enhancement deviations with respect to classical, local model.

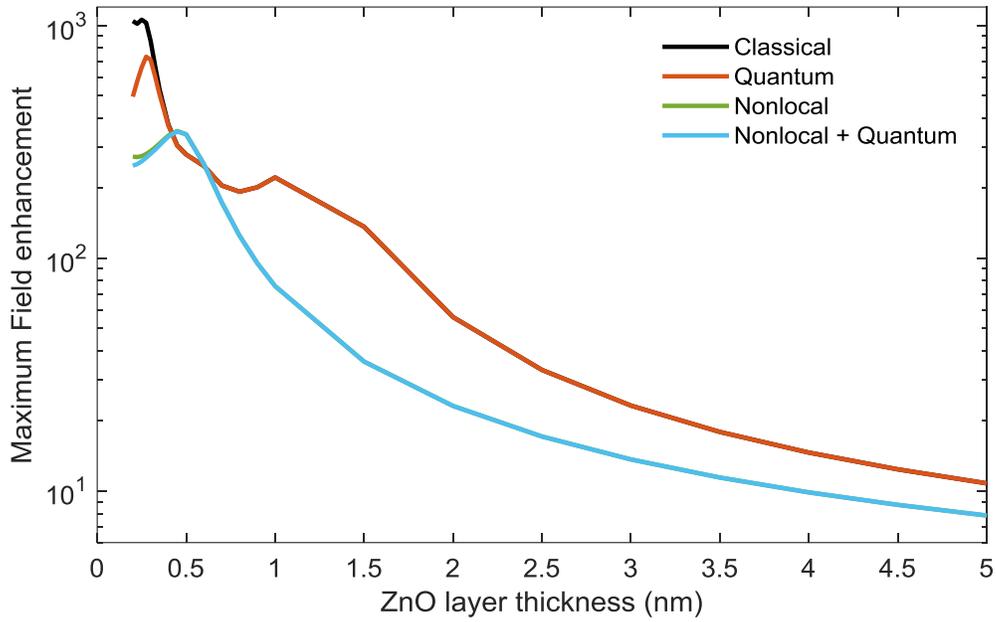

Fig. S9. Maximum field enhancement in the gap region of an MIM structure as a function of gap size for the four cases described in the text.

In Fig. S10 the distribution of the field enhancement in the gap region is plotted for case (iv) when the ZnO thickness is only 0.2 nm. Besides setting a limit to the maximum field enhancement (~250 in this scenario), the effect of the nonlocality is observable as a substantial field penetration in the metallic regions.

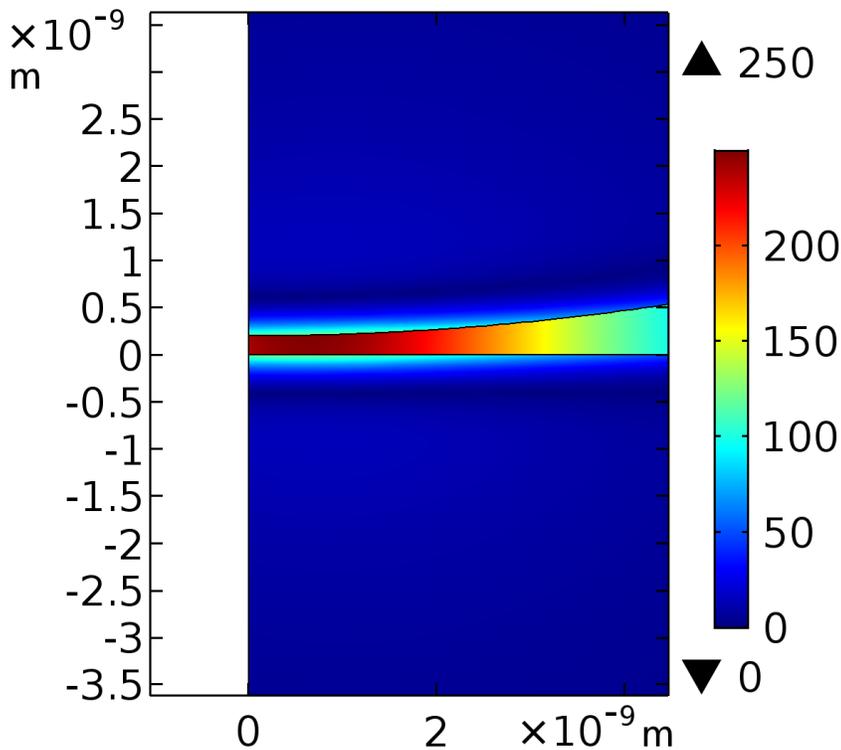

Fig. S10. Field enhancement distribution near the gap region for an MIM with a gap size of 0.2 nm, i.e., first point data in Fig. S9. Here both quantum and nonlocal effects are considered in the simulation.

**REFERENCES**

1. P. K. Tien, and J. P. Gordon, "Multiphoton Process Observed in the Interaction of Microwave Fields with the Tunneling Between Superconductor Films," Phys. Rev. **129**, 647 (1963).
2. J. R. Tucker, and M. F. Millea, "Photon detection in nonlinear tunneling devices," Appl. Phys. Lett. **33**, 611 (1978).
3. J. R. Tucker, and M. J. Feldman, "Quantum detection at millimeter wavelengths," Rev. Mod. Phys. **57**, 1055 (1985).
4. J. W. Haus, D. de Ceglia, M. A. Vincenti, and M. Scalora, "Quantum Conductivity for Metal-Insulator-Metal Nanostructures," J. Opt. Soc. Am. B **31**, 259 (2014). arXiv 1309.1363.
5. J. W. Haus, D. de Ceglia, M. A. Vincenti, and M. Scalora, "Nonlinear quantum tunneling effects in nano-plasmonic environments," J. Opt. Soc. Am. B **31**, A13-A19 (2014)
6. J. G. Simmons, "Generalized formula for the electric tunnel effect between similar electrodes separated by a thin insulating film," J. Appl. Phys. **34**, 1793 (1963).
7. J. G. Simmons, "Electric tunnel effect between dissimilar electrodes separated by a thin insulating film," J. Appl. Phys. **34**, 2581 (1963).
8. P. E. Powers and J. W. Haus, Fundamentals of Nonlinear Optics 2nd Ed., (CRC Press, 2017).
9. Z. Gao, M. M. R. Hussain, D. de Ceglia, M. A. Vincenti, A. Sarangan, I. Agha, M. Scalora, J. W. Haus, and P. Banerjee, "Unraveling delocalized electrons in metal induced gap states from second harmonics," Appl. Phys. Lett. **111**, 161601 (2017).
10. J. F. Moulder, W. F. Stickle, P. E. Sobol, and K. D. Bomben, , *Handbook of X-ray photoelectron spectroscopy: a reference book of standard spectra for identification and interpretation of XPS data*, Physical Electronics: Eden Prairie, MN (1995).
11. B. V. Crist, *Handbooks of monochromatic XPS spectra, Vol. 1*, XPS International Mountain View, CA (1999).
12. R. H. Ritchie, "Plasma losses by fast electrons in thin films," Phys. Rev. **106,** 874 (1957)
13. C. Ciracì, Y. Urzhumov, and D. Smith, "Far-field analysis of axially symmetric three-dimensional directional cloaks," Opt. Express **21**, 9397 (2013).